\documentstyle[twoside,fleqn,espcrc2,epsfig]{article}

\newcommand{\AmS}{{\protect\the\textfont2
  A\kern-.1667em\lower.5ex\hbox{M}\kern-.125emS}}
\newcommand{\Ns}{{N_{\sigma}}}
\newcommand{\Nt}{{N_{\tau}}}

\newcommand{\tr}{{\rm Tr~}}
\newcommand{\bq}{\begin{equation}}  
\newcommand{\eq}{\end{equation}}
\newcommand{\bqa}{\begin{eqnarray}}
\newcommand{\eqa}{\end{eqnarray}}

\title{ Finite Temperature Phase Diagramm of QCD with improved Wilson fermions}

\author{ M. Oevers with F. Karsch \thanks{Work supported by NATO grant
    no. CRG940451}, E. Laermann, P. Schmidt\address{ Fakult{\"a}t f{\"u}r 
    Physik, 
    Universit{\"a}t Bielefeld, Postfach 100131, 33501 Bielefeld, Germany}}
       
\begin{document}
\pagestyle{empty}
\begin{abstract}
  We present first results of a study of two flavour QCD with
  Wilson fermions at finite temperature. We have used tree 
  level Symanzik improvement in both the gauge and fermion 
  part of the action. In a first step we explore the phase
  diagramm on an $8^3 \times  4$ lattice, with particular 
  emphasis on checking Aoki's conjecture with an improved 
  action.
\end{abstract}

\maketitle

\section{Introduction}
The study of the finite temperature phase transition in QCD with Wilson 
fermions is much more complicated than in the staggered fermion formulation,
because of the absence of an order parameter due to explicit chiral symmetry
breaking. This means e.g. that the existence of a massless pion at finite
lattice spacing is not at all obvious.

In recent years a detailed picture of the finite temperature phase diagramm of
QCD with Wilson fermions has emerged \cite{Phdiag}. This picture is based on 
the idea of spontaneous breakdown of parity and flavour symmetry \cite{Aoki-I} 
and has been investigated analytically as well as numerically.
The features of this phase diagramm are 
(i) the critical line $\kappa_c(\beta)$ defined by a vanishing pion screening
mass for finite  temporal  lattice size turns back towards strong coupling
forming a cusp,
(ii) the region bounded by the critical line represent a phase of spontaneously
broken parity and flavour symmetry, 
(iii) the finite temperature phase transition line $\kappa_t(\beta)$ presumably
does not cross the critical line, but runs past it towards larger values of 
the hopping parameter\cite{Aoki-II}.

From an analysis of the Gross-Neveu model in two dimensions, where three cusps
connected to doublers develop, one expects the critical line for QCD in four 
dimensions to form five cusps moving towards weak coupling with increasing
temporal lattice size $N_\tau$. Simulations with the standard Wilson 
formulation for quarks and gluons have shown, that at  $N_\tau=1/(aT)=4$ the 
tip of the cusp lies in the strong coupling regime at $\beta  \approx 3.9$ and
moves only slowly towards weak coupling as $N_\tau$ is increased. 

Since one
expects the same features to hold for a wider class of actions including the
clover action, which tends to reduce cutoff dependencies, a study of
the phase diagramm using improved actions is of practical as well as 
theoretical importance\cite{improv}.

\section{Results}
We have conducted a simulation of 2 flavour QCD on an $8^3 \times 4$ lattice 
using tree level Symanzik improved actions for quarks and gluons. In the gauge
sector this amounts to adding a $2 \times 1$-loop to the standard plaquette
action and for the fermions in adding the clover term with $c_{SW}=1$.
We have used
a Hybrid Monte Carlo algorithm with a timestep $\delta\tau=0.01$ and the number
of molecular dynamics steps $N_{MD}=20$, which so far amounts to rather short
trajectories. We simulated several $\kappa$ values 
for each  $\beta=3.00, 3.50, 3.75$ and $4.00$. We have measured the Polyakov
loop, the pion norm and the average number of iterations it takes to invert 
the fermion matrix. Each observable is now discussed in detail.

\subsection{Locating the critical line}
Although not a physical observable in its own right, the average number of
iterations it takes to invert the fermion matrix is a very good indicator for
criticality. 
The simulations where done in the following way. At each $\beta$-value we
started a simulation at $\kappa=0.12$ and used a thermalized configuration from
this run as a start configuration at a higher value of $\kappa$. We continued 
to do so for higher and higher $\kappa$-values. At those $\beta$-values, where
we saw a drastic increase in the number of iterations, we started a simulation
at a much higher $\kappa$-value and continued towards smaller $\kappa$-values.
Our findings are summarized in {Fig.\,1}. At $\beta=3.00$ we were not
able to simulate the system for $0.1770<\kappa<0.1825$. At $\beta=3.50$ we
saw an initial increase of iterations, even after switching from minimal 
residual to conjugate gradient, which usually decreased the number of 
iterations. With increasing $\kappa$ the number of iterations decreased
again, only to increase again at fairly high $\kappa$-values. At $\beta=3.75$
and $4.00$ we saw a similar behaviour as at $\beta=3.50$ but not as pronounced.
We experimented with different inversion routines and our conclusions are, that
close to the critical line conjugate gradient is superiour to overrelaxed
minimal residual and BiCGstab1.
\begin{figure}[t]
  \begin{center}
    \epsfig{file=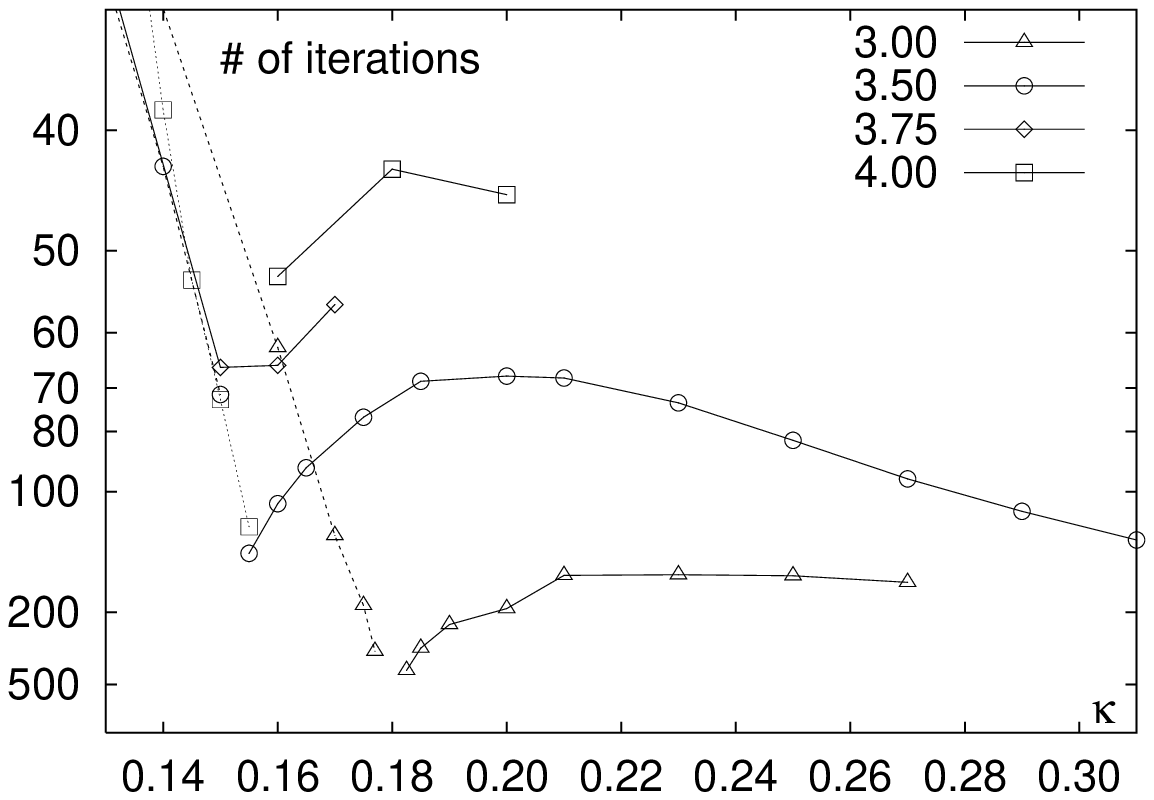, width=7.75cm}
  \end{center}
    {\small Fig.\,1 Average number of iterations as a function of
      $\kappa$ for $\beta=3.00$ (triangles), $\beta=3.50$ (circles),
      $\beta=3.75$ (diamonds) and $\beta=4.00$ (boxes). The inverter used are
      conjugate gradient (lines), overrelaxed minimal residual (dashed),
      BiCGstab1 (dotted)}  
\end{figure}

\subsection{Pion Norm}
Since it is not possible to reliably extract the pion screening mass on small 
lattices, we use the pion norm instead to indicate the existence of a critical
line of vanishing pion screening mass.
The pion norm is the integrated pion correlator and is defined as
follows:
\bq
\Pi= \frac{1}{4 \Ns^3 \Nt} \cdot \tr 
    \left[{\cal M}^{-1} \gamma_5 {\cal M}^{-1} \gamma_5\right]
\eq
Here ${\cal M}$ is the fermion matrix on a particular gauge configuration.
Near the critical line the pion norm behaves as $\Pi \approx 1/m_{\pi}^2$, 
hence a diverging pion norm indicates the existence of the critical line.
Our results are displayed in {Fig.\,2}. At $\beta=3.00$ we find a clear signal 
for two critical lines close to $\kappa=0.1770$ and $0.1825$. 
The difference in 
$\kappa$ is already quite small, so we are near the tip of the cusp. At
$\beta=3.50$ the pion norm develops a small peak at $\kappa=0.1550$, which is
located where the crossover from the low temperature to the high temperature
phase starts (see section on Polyakov loop). No divergent behaviour is observed
and the critical line ceases to exist in this coupling region. The simulations
at $\beta=3.75$ and $4.00$ do  not even find a peak for the pion norm, but
rather a smooth behaviour as a function of hopping parameter. We conclude, that
the critical line turns back towards strong coupling and develops a cusp 
between $\beta=3.00$ and $3.50$. 
  
\begin{figure}[t]
  \begin{center}
    \epsfig{file=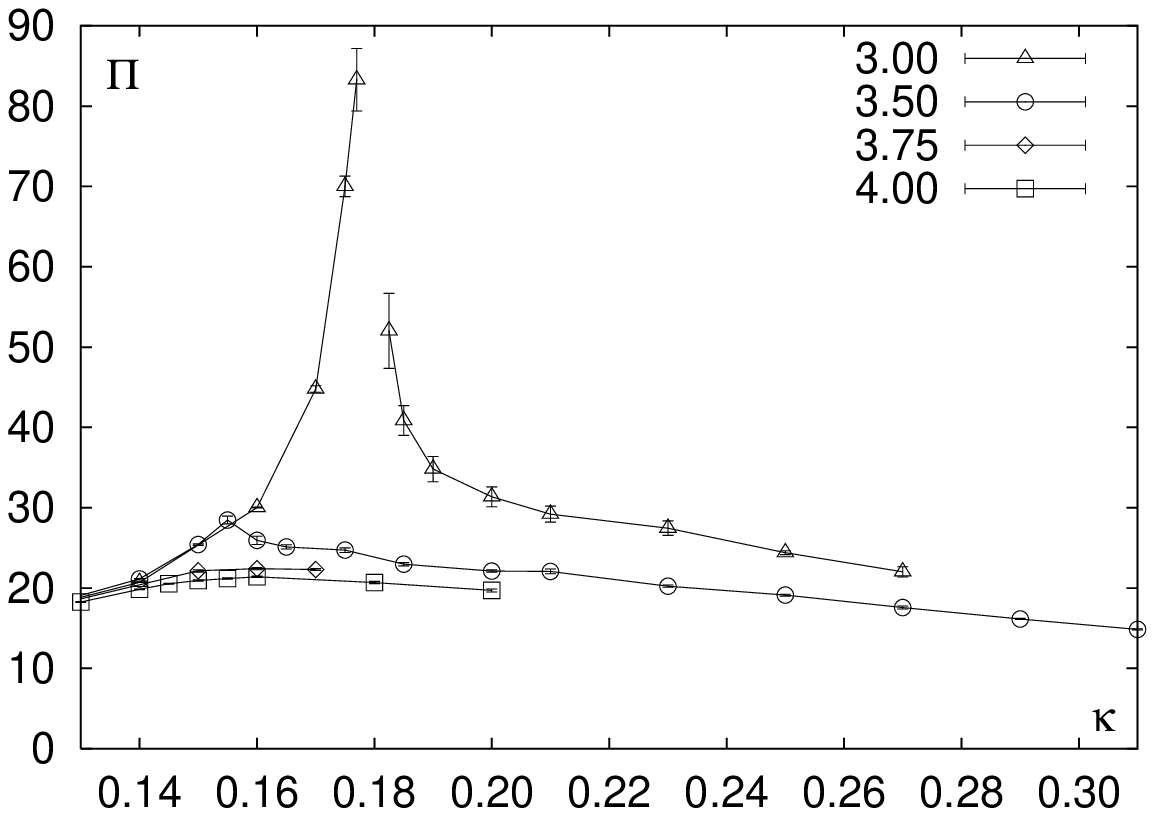, width=7.75cm}
  \end{center}
    {\small Fig.\,2 Pion norm as a function of $\kappa$ for $\beta=3.00$
      (triangles), $\beta=3.50$ (circles), $\beta=3.75$ (diamonds) and
      $\beta=4.00$ (boxes).}
  \vspace{-1ex}
\end{figure}

\subsection{Polyakov Loop}
The Polyakov loop is defined as follows:
\bq
    L = \frac{1}{\Ns^3} \sum_{\vec{x}} \frac{1}{N_c} \tr \prod_{\tau=1}^{\Nt} 
        U_4(\vec{x},\tau)
\eq
This observable is sensitive to the finite temperature phase transition 
although it is no order parameter in the full theory. Our results are displayed
in {Fig.\,3}. At $\beta=3.00$ we find
a confined phase for $\kappa \le 0.1770$. For for $\kappa > 0.1825$, when one 
approaches the critical line from above, the  Polyakov loop decreases to
$|L| \approx 0.1$. This indicates that the system develops a low temperature 
behaviour in the vicinity of the critical line. On the other hand we do
not see a sharp crossover, so we cannot conclude that one crosses the thermal
line as one lowers $\kappa$ towards $\kappa_c(\beta)$.

At $\beta=3.50$ the Polyakov loop displays a sharp crossover phenomenon, which
means that the system crosses the thermal line for 
{$0.1550 < \kappa < 0.1600$}. 

At $\beta=3.75$ and $4.00$ the system is in the high temperature phase down to 
$\kappa=0.12$. This is not unexpected, since the finite temperature phase
transition in the quenched theory for our choice of action occurs at 
$\beta_c=4.07$. 
\begin{figure}[t]
  \begin{center}
    \epsfig{file=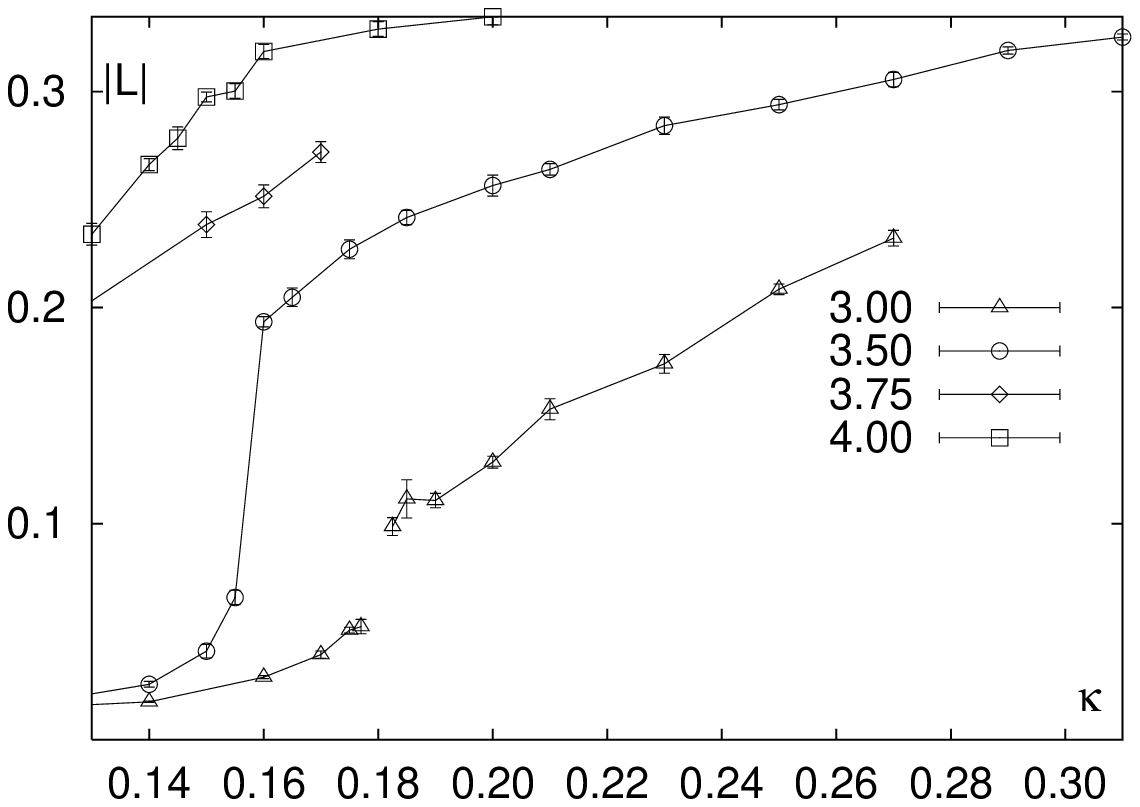, width=7.75cm}
  \end{center}
    {\small Fig.\,3 Polyakov loop as a function of $\kappa$ for
      $\beta=3.00$ (triangles), $\beta=3.50$ (circles), $\beta=3.75$ (diamonds)
      and $\beta=4.00$ (boxes).}

\end{figure}
\section{Summary and Conclusions}
From measuring the pion norm, Polyakov loop and the average number of 
iterations to invert the fermion matrix we conlude, that the finite temperature
phase structure observed with the standard Wilson formulation is preserved for
the tree level Symanzik improved formulation of quarks and gluons. We note that
the difference in $\beta$ between the location of the cusp and the quenched
$\beta_c$ is considerably reduced. What this means in terms of physical scales
remains to be investigated by measurements of the lattice spacing. After this
preparatory simulation on a small lattice, we will investigate the phase 
diagramm on larger lattices including a precise measurement of the
pion screening mass and  quark mass, the latter enabling us to study the chiral
condensate as well \cite{ChWI}.

\end{document}